\documentclass[review]{elsarticle}

\usepackage{lineno,hyperref}
\usepackage{amsmath}

\journal{Journal}

\begin{document}
\begin{frontmatter}

\title{Signal fluctuations and the Information Transmission Rates in binary communication channels}

\author{Agnieszka Pregowska}
\address{Institute of Fundamental Technological Research, Polish Academy of Sciences, Pawinskiego 5B, 02–106
	Warsaw, Poland}




\begin{abstract}
\par
In nervous system information is conveyed by a sequence of action potentials, called spikes-trains, which can be represented as bits coming from Information Sources $(IS)$. Previously, we studied relations between spikes Information Transmission Rates $(ITR)$, their correlations, and frequencies. Here, we concentrate on the problem of how spikes fluctuations affect $ITR$.
\par
$IS$ are modeled as stationary stochastic processes, which we assume here as two-state Markov processes. As a spike-trains' fluctuation measure, we consider the standard deviation $\sigma$, which measures average fluctuation of spikes around average spike frequency. 
\par
We found that character of $ITR$ and signal fluctuations relation strongly depends on parameter $s$ being a sum of transitions probabilities from no spike to spike states. It turned out that for $s<1$ the quotient $\frac{ITR}{\sigma}$ has a maximum and can tend to zero depending on transition probabilities, while for $1<s \frac{ITR}{\sigma}$ is separated from 0. An estimate of $ITR$ was found by expressions depending on: signal fluctuation, $s$ parameter and entropy of  corresponding Bernoulli process.
\par
Our results show that in a more noisy environment, to get appropriate reliability and efficiency of transmission, $IS$ with higher tendency of transition from the state no spike to spike state and \textit{vice versa} should be applied.

\end{abstract}

\begin{keyword}
information source \sep information transmission rate \sep fluctuations \sep Shannon entropy \sep standard deviation \sep spike-trains.
\end{keyword}

\end{frontmatter}


\section{Introduction}
\par
Information transmission processes in natural environments are usually affected by signals fluctuations due to the presence of noise-generating factors \cite{Weber2020, HukHart2019}. It is especially visible in biological systems, in particular in signal processing in the brain \cite{vanHemmen2006, Deco2009, Fraiman2012,Nazari2019,Nazari2019a}. The physical information carriers in the brain are small electrical currents \cite{Adrian1926}. Specifically, the information is carried by sequences of action potentials also called spikes-trains. Assuming some time resolution MacKay and McCulloch proposed a natural encoding method that associates to each spike-train a binary sequence \cite{MacKay1952}. Thus, the information is represented by a sequence of bits which, from a mathematical point of view, can be treated as a trajectory of some stochastic process \cite{Cover1991, Rieke1997}. 
\par
In 1948 C. Shannon developed his famous Communication Theory where he introduced the concept of information and its quantitative measure \cite{Shannon1948}. The occurrences of both inputs transmitted through a communications channel and output symbols are described by sequences of random variables that define already stochastic processes and form some Information Sources \cite{Cover1991,Ash1965}. Following this line, to characterize the amount of information transmitted per symbol the Information Transmission Rate $(ITR)$ is applied. 
\par
Spike-trains Information Sources are often modeled as Poisson point processes \cite{Teich1985,Daley2003}. On the other hand, it is known that such processes exhibit Markov properties \cite{Ross1996,Papoulis2002}. This is because in these processes when describing spikes arrival times, current time and the time from the last spike is primarily taken into account \cite{Kass2001}. 
\par
Description of complex systems dynamics, from financial markets \cite{Bouchaud2004,Iovane2016,Sang2019} to the neural networks of living beings \cite{Knoblauch2005,Mishkovski2011}, require appropriate mathematical tools. Among them there are stochastic processes, Information Theory and statistical methods and recently, fuzzy numbers \cite{Zadeh1965,Prokopowicz2019}. Traditionally, the complex nature of systems is characterized, mostly due to the presence of noise, by using  fluctuations, variations, or other statistical tools \cite{Zadeh1965}. The natural measure of fluctuations should, in general, reflect oscillations around the mean average value of the signal. Therefore, in most systems in physics, economics, fluid mechanics, fluctuations are most often quantifying using the Standard Deviation \cite{Frisch1995,Salinas2000,Kittel2004}. 
\par
In this paper, we analyze the relationship between the Information Transmission Rate of signals coming from time-discrete two states Markov Information Source and these signals fluctuations. As a spike-trains' fluctuation measure, we consider already the Standard Deviation of encoded spikes. Moreover, to get the better insight we have also analyzed the case when the $ITR$ is referred to the signals Variance $V$ instead to the Standard Deviation $\sigma$. 
\par
Our previous research, when we studied the properties of neural coding, shows that neural binary coding cannot be captured by straightforward correlations between input and output signals \cite{Pregowska2015}. In \cite{Pregowska2016,Pregowska2019} it was found that a key role in assessing the information sent by Markov type Information Sources in dependence on both Firing Rate and signals correlations plays the jumping (transition) parameter $s$, which is the sum of transition probabilities from the no-spike state to the spike state and \textit{vice versa}. Here, we found that also the character of the relation between I$TR$ and signal fluctuations strongly depends on the parameter $s$. It turned out that for small $s$ $(s<1)$ the quotient $\frac{ITR}{\sigma}$ has a maximum and tends to zero when the probability of transition from no spike state to spike state never reaches 0. While for large enough $s$ the quotient $\frac{ITR}{\sigma}$ is limited from below. We observed that similar behavior of $\frac{ITR}{\sigma}$  is also when we replaced (approximate) Shannon entropy formula by appropriate polynomials. 
\par
On the other hand, we found that when we refer the quotient $\frac{ITR}{\sigma}$ to $\sigma$, i.e. when we consider, in fact, the quotient $\frac{ITR}{V}$ this quotient behaves in a completely different way. This behavior is not regular. Specifically, we observed that for $1<s$ there is some range of parameter $s$ for which $\frac{ITR}{V}$ has a few local extremas, in opposition to the case $\frac{ITR}{\sigma}$.  
\par
The paper is organized as follows. In Section \ref{theory}, we briefly recall Shannon Information Theory concepts (entropy, information, binary Information Sources, Information Transmission Rate), and fluctuation measure (Standard Deviation and Root Mean Square). In Section \ref{results} we analyzed the quotients $\frac{ITR}{\sigma}$ and $\frac{ITR}{V}$. Section \ref{disscussion} contains the discussion and final conclusions.
\section{Theoretical Background and Methods}\label{theory}
To introduce the necessary notation, we briefly recall Shannon Information Theory's basic concepts \cite{Cover1991,Shannon1948,Ash1965} i.e. Information, Entropy, Information Source, and Information Transmission Rate. 
\subsection{Shannon’s Entropy and Information Transmission Rate}
Let $Z^{L}$ be a set of all words of length $L$, built of symbols (letters) from some finite alphabet $Z$. Each word $w \in Z^{L}$ can be treated as an encoded message sent by Information Source $\textbf{Z}$ being a stationary stochastic process. If $P(w)$ denotes the probability that the word $w \in Z^{L}$ already occurs, then the information in the Shannon sense carried by this word is defined as
\begin{equation}\label{Shannon_Information}
I(w):=-\log_{2}{P(w)}  .
\end{equation}
This means that less probable events carry more information. Thus, the average information of the random variable $\textbf{Z}^{L}$ associated with the words of length $L$ is called the Shannon block entropy and is given by 
\begin{equation}\label{Shannon_Block_Entropy}
H(\textbf{Z}^{L}):=-\sum_{w \in Z^{L}}P(w)\log_{2} P(w) .
\end{equation} 
The appropriate measure for estimation of transmission efficiency of an Information Source $\textbf{Z}$ is the information transmitted on average by a single symbol, i.e. $ITR$ \cite{Cover1991,Ash1965} 
\begin{equation}\label{itr}
ITR^{(L)}(\textbf{Z}):=\frac{1}{L}H(\textbf{Z}^{L})
\end{equation}
\begin{equation}\label{itr_2}
ITR(\textbf{Z})=  \lim_{L \rightarrow \infty}\frac{1}{L}H(\textbf{Z}^{L})  .
\end{equation}
This limit exists if and only if the stochastic process $\textbf{Z}$ is stationary \cite{Cover1991}.
\par
In the special case of a two-letters alphabet $Z=\{0,1\}$ and the length of words $L=1$ we introduce the following notation
\begin{equation}\label{entropy_bin}
H_{2}(p):=H(\textbf{Z}^{1})=-p \log_{2}p-(1-p) \log_{2}(1-p) .
\end{equation} 
where $P(1)=p,P(0)=1-p$ are associated probabilities. This is, in fact, the formula for the entropy rate of a Bernoulli source \cite{Ash1965}. Index 2 in (\ref{entropy_bin}) indicates that we consider logarithm wit base 2 what means that we consider the information expressed in bits. 
\subsection{Information Sources}\label{IS}
In general, Information Sources are modeled as stationary stochastic processes \cite{Cover1991,Ash1965}. The information is represented by trajectories of such processes. Here, to study the relation between Information Transmission Rate $(ITR)$ and trajectories fluctuations, we consider Information Sources which are modeled as two-states Markov processes. The trajectories of these processes can be treated as encoded spike-trains \cite{vanHemmen2006,Rieke1997,Amigo2004}. The commonly accepted natural encoding procedure leads to binary sequences \cite{Rieke1997,Amigo2004}. Spike-trains are, in fact, the main objects to carry information \cite{vanHemmen2006,Adrian1926}. We additionally consider among the Markov processes as a special case the Bernoulli processes.
\subsubsection{Information Sources – Markov Processes}
We consider time-discrete, two-states Markov process $\textbf{M}$, which is defined by a set of conditional probabilities $p_{j|i}$ which describe the transition from state $i$ to state $j$, where $i,j=0,1,$ and by the initial probabilities $P_{0} (0),P_{0} (1)$. The Markov transition probability matrix $\textbf{P}$ can be written as
\begin{equation}\label{Markov_tras_prob}
\textbf{P}:= 
\begin{bmatrix}
p_{0|0} & p_{0|1}  \\
p_{1|0} & p_{1|1} 
\end{bmatrix}
=
\begin{bmatrix}
1-p_{1|0} & p_{0|1}  \\
p_{1|0} & 1-p_{0|1} 
\end{bmatrix}
.
\end{equation}
Each of the columns of the transition probability matrix $\textbf{P}$ has to sum to 1 (i.e. it is a stochastic matrix \cite{Cover1991}). 
\par
The time evolution of the states probabilities is governed by the Master Equation \cite{vanKampen2007}
\begin{equation}\label{Master_Equation}
\begin{bmatrix}
P_{n+1}(0) \\ P_{n+1}(1)
\end{bmatrix}
=
\begin{bmatrix}
1-p_{1|0} & p_{0|1}  \\
p_{1|0} & 1-p_{0|1} 
\end{bmatrix}
\cdot
\begin{bmatrix}
P_{n}(0) \\ P_{n}(1)
\end{bmatrix}
\end{equation}
where $n$ stands for time, $P_{n}(0),P_{n}(1)$ are probabilities of finding states $"0"$ and $"1"$ at time $n$, respectively. The stationary solution of (\ref{Master_Equation}) is given by
\begin{equation}\label{stacionary_solution}
\begin{bmatrix}
P_{eq}(0) \\
P_{eq}(1)
\end{bmatrix}
=
\begin{bmatrix}
\frac{p_{0|1}}{(p_{0|1}+p_{1|0})} \\
\frac{p_{1|0}}{(p_{0|1}+p_{1|0})}
\end{bmatrix}
.
\end{equation}
It is known \cite{Cover1991,Ash1965} that for Markov process $\textbf{M}$ the Information Transmission Rate as defined by (\ref{itr_2}) is of the following form
\begin{equation}\label{entropy_Markov}
H^{\bf{M}}=P_{eq}(0) \cdot H(p_{1|0})+P_{eq}(1) \cdot H(p_{0|1})
\end{equation}
In previous papers \cite{Pregowska2015,Pregowska2016,Pregowska2019}, when we studied the relation between $ITRs$ and firing rates and when we compared $ITR$ for Markov processes and for corresponding Bernoulli processes we have introduced a parameter $s$, which can be interpreted as the tendency of a transition from the no-spike state $("0")$ to the spike state $("1")$ and \textit{vice versa}:
\begin{equation}\label{s}
s:=p_{0|1}+p_{1|0}
\end{equation}
It turned out that this parameter plays an essential role in our considerations also in this paper. Note that $s=2-tr\bf{M}$ and $0 \leq s \leq 2$. One can observe that two-states Markov processes are Bernoulli processes if and only if $s=1$. 
\subsubsection{Information Sources – Bernoulli Process case}
The Bernoulli processes play a special role among the Markov processes. Bernoulli process is a stochastic stationary process $\bf{Z}$$=(Z_i ),i=1,2,… $ formed by binary identically distributed and independent random variables $Z_{i}$. In the case of the encoded spike-trains, we assume that the corresponding process (to be more precise its trajectories) takes successively the values 1 (when spike has arrived in the bin) or 0 (when spike has not arrived). We assume that for a given size of time-bin applied (this depends in turn on the time resolution assumed), spike trains are encoded \cite{Bialek1991} in such a way that 1 is generated with probability $p$, and 0 is generated with probability $q$, where $q$ is equal to $1-p$. Following the definition, the Information Transmission Rate (\ref{itr}) of the Bernoulli process is
\begin{equation}\label{entropy_Bernoulli}
H^{B}(p,q)=-p \log_{2}p-q \log q= H_{2}(p).
\end{equation}
\subsubsection{Generalized entropy variants}
\begin{figure}
	\centering
	\includegraphics[width=12 cm]{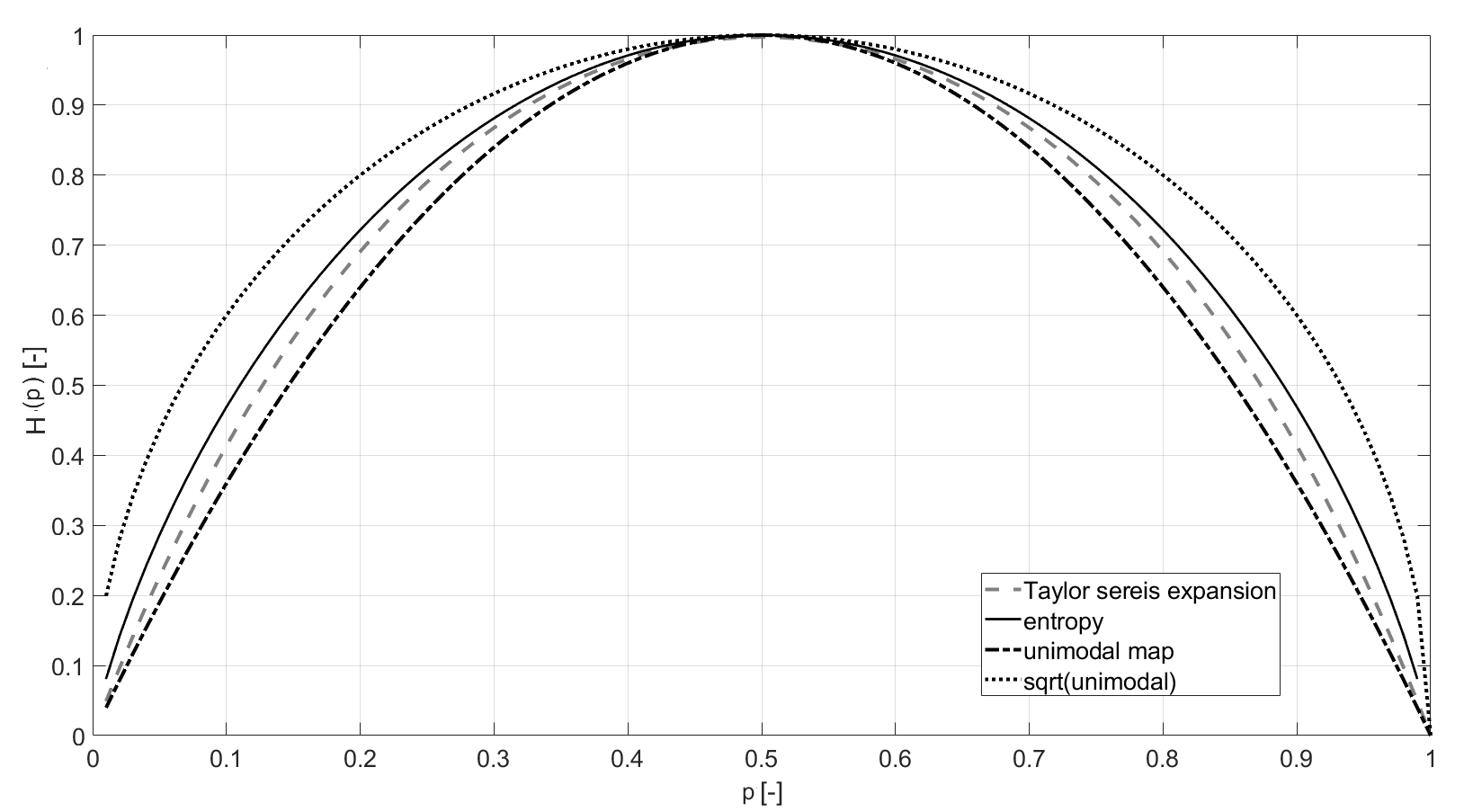}
	\caption{Approximation of the Shannon entropy (black solid lines) using the Taylor series expression (gray dash-dot line, 10 first terms), unimodal function (black dash line), and unimodal map root (black point line).} \label{function}
\end{figure}

The form of entropy $H$ was derived under assumptions of monotonicity, joint entropy, continuity properties, and Grouping Axiom. In the classical case of the entropy rate $H^{\bf{M}}$  for Markov process, in formula  (\ref{entropy_Markov}) the terms $H(p_{1|0})$ and $H(p_{0|1})$ are clearly understood in the Shannon sense (\ref{Shannon_Block_Entropy}). To get a better insight into the asymptotic behavior of the relations studied in this paper, we additionally consider formula (\ref{entropy_Markov}) with $H$ replaced by its Taylor approximation (10 terms). We also studied the interesting case when instead of H we used famous unimodal map $U(p)=4p(1-p)$ \cite{ColletEckmann1980} which is, in fact, close (Figure \ref{function}) to $H$ in the supremum norm \cite{Rudin1964}. This idea is along the research direction related to generalized concepts of entropy developed, starting from Renyai \cite{Renyi1960}, by many authors \cite{Amigo2010,Crumiller2013,Bossomaier2016,Amigo2018,Jetka2018}. Figure \ref{function} shows the approximation of entropy (\ref{entropy_Markov}) by polynomials: unimodal map (black dash line) and 10 first terms in the Taylor series of $H$ (gray dash-dot line). We also included the square root of the unimodal map (black point line) in this Figure.
\subsection{Fluctuations measure}
It is commonly accepted that for a given random variable $\bf{X}$ the fluctuations of values of this random variable around its average can be characterized by the Standard Deviation $\sigma$ \cite{Feller1958}
\begin{equation}\label{standard_deviation}
\sigma :=(\bf{E}(\bf{X}-E\bf{X})^{2})^{\frac{1}{2}}
\end{equation}
where symbol $\bf{E}$ means the average taken over the probability distribution associated with the values reached by $\bf{X}$.
\par
Considering a stochastic process $\bf{Y}$$=(X_{k}), k=1,2,3, \ldots $, where $X_{k}$ are random variables each with the same probability distribution as $\textbf{X}$, the fluctuation of trajectories of this process can be estimated by the Root-Mean-Square $(RMS)$. For a given trajectory $(x_{k})_{i=1}^{n},k=1, \ldots,n$ RMS is defined as the root from the arithmetic mean value of the squares, i.e.

\begin{equation}\label{rms}
RMS(\textbf{Y}):=(\frac{1}{n} \sum_{k=1}^{n}(x_{k}-x_{n_{avr}})^{2})^{\frac{1}{2}}
\end{equation}
where  $x_{n_{avr}}$ is the average value, i.e. $x_{n_{avr}}=\frac{1}{n} \sum_{i=1}^{n}x_{k}$. Note, that from this formula the form of $\sigma$ for Markov processes can be derived when using stationary distribution (\ref{stacionary_solution}) in formula (\ref{standard_deviation}).
\par
The Standard Deviation $\sigma$ for any random variable depends, in fact, not only on its probability distribution but also on the values taken by this random variable. Here we are interested in bits oscillation, i.e. if the spike train occurs or not. Thus, we have limited our considerations to the values 0 and 1.
\par
To get a better insight into the relation between $ITR$ and $signal/bits$ fluctuations we also included an analysis of the quotient $\frac{ITR}{V}$. This is interesting due to the specific form of Variation for the Bernoulli process what leads to interesting observations when consider, for example, the unimodal map to approximate entropy (\ref{entropy_bin}). Moreover, when studying $\frac{ITR}{V}$ we, in fact, refer the quotient $\frac{ITR}{\sigma}$ to $\sigma$ since we have simply $\frac{(\frac{ITR}{\sigma})}{\sigma}$=$\frac{ITR}{V}$.
%
%
%
%
\section{Results}\label{results}
In this Section, we study the quotients  $\frac{ITR}{\sigma}$ and $\frac{ITR}{V}$ as a function of the transition probability $p_{1|0}$ from the state no-spike $"0"$ to the spike state $"1"$ for a fixed parameter $s$ (\ref{s}). Note, that the probability $0<p_{1|0}<1$ and parameter  $0<s<2$ uniquely determined the transition probability matrix $\bf{P}$ (\ref{Markov_tras_prob}) and consequently, they completely define the Markov process $\bf{M}$, provided that initial probabilities $P_{0}(0),P_{0}(1)$ are chosen. Here, as initial probabilities, to get a stationary process, we must assume the probabilities of the form (\ref{stacionary_solution}). 
\subsection{Information against fluctuations for two-states Markov processes – general case}
We start our considerations from the most general form of the two-states Markov process. To analyze the quotients $\frac{ITR}{\sigma}$ and $\frac{ITR}{V}$ we first express Standard Deviation of Markov process $\bf{M}$ in terms of conditional probability $p_{1|0}$ and parameter $s$.
\subsubsection{Standard Deviation in the Markov process case}\label{MarkovCase}
For a given Markov process $\bf{M}$ to evaluate its fluctuation, specifically to address its long time  behavior, one considers its corresponding stationary probabilities as defined by (\ref{stacionary_solution}). Thus, in the limiting case, the Standard Deviation $\sigma$ for the Markov process can be assumed as 
\begin{equation}\label{standard_deviation_Markov}
\sigma^{\bf{M}}=\sqrt{P_{eq}(0) \cdot P_{eq}(1)} .
\end{equation}
Fixing parameter $s$ and expressing $\sigma^{\bf{M}}$ as a function of the conditional probability $p_{1|0}$ we came to the following formula:
\begin{equation}\label{standard_deviation_Markov_s}
\sigma_{s}^{\bf{M}}(p_{1|0})=\sqrt{\frac{p_{0|1}}{s} \cdot \frac{p_{1|0}}{s}}=\frac{\sqrt{(s-p_{1|0})p_{1|0}}}{s} .
\end{equation}
Note that in the case of Variance $[V_{s}^{\bf{M}}(p_{1|0})]=[\sigma_{s}^{\bf{M}}(p_{1|0})]^{2}$ we have a polynomial dependence on $p_{1|0}$ (keeping in mind that $s$ is fixed).
\subsubsection{Relation between Information Transmission Rate $ITR$ of Markov process and its Standard Deviation }
Let’s start by establishing the relation between Standard Deviation and $ITR$ for the Bernoulli process. This means that in our notation $s$ is equal to 1. Making use of the classical inequality $x-1 \geq \ln{x}$$($for all $x>0$$)$ and doing a few simple operations one can come to the inequality $2 \cdot \log_{2}e \leq \frac{ITR_{2}(p_{1|0})}{\sigma^{2}}$.
To find the relations between Information Transmission Rate $ITR^{\bf{M}}$ and $\sigma^{\bf{M}}$ in more general cases, one can consider the quotient 
\begin{equation}\label{Q_M}
Q_{\sigma}^{\bf{M},s}(p_{1|0}):=\frac{ITR^{\bf{M},s}(p_{1|0})}{\sigma^{\bf{M}}_{s}(p_{1|0})} .
\end{equation}
Note that $Q_{\sigma}^{\bf{M},s}(p_{1|0})$ is a symmetric function with respect to to the axe $p_{1|0}=\frac{s}{2}$ i.e. 
\begin{equation}\label{Q_M_symmetry}
Q_{\sigma}^{\bf{M},s}(p_{1|0})=Q_{\sigma}^{\bf{M},s}(s-p_{1|0}) .
\end{equation}
For $0 \leq s \leq 2$, we consider the quotient $Q_{\sigma}^{\bf{M},s}(p_{1|0})$ in two cases taking into account the range of $p_{1|0}$
\begin{equation}\label{A}
\mbox{A)} \quad 0\leq s \leq 1 \quad \mbox{and this implies } \quad 0 \leq p_{1|0} \leq s
\end{equation}
\begin{equation}\label{B}
\mbox{B)} \quad 1 < s <2 \quad \mbox{and this implies} \quad s-1 \leq p_{1|0} \leq 1.
\end{equation}
Substituting (\ref{stacionary_solution}), (\ref{s}) and (\ref{standard_deviation_Markov}) into (\ref{Q_M}) we obtain
\begin{equation}\label{Qsigma}
Q_{\sigma}^{\bf{M},s}(p_{1|0}):=\frac{\frac{p_{0|1}}{s}H(p_{1|0})+\frac{p_{1|0}}{s}H(p_{0|1})}{\frac{\sqrt{(s-p_{1|0})p_{1|0}}}{s}}
\end{equation}
and after simple calculations we have
\begin{equation}
\begin{split}
Q_{\sigma}^{\bf{M},s}(p_{1|0})=\frac{\frac{s-p_{1|0}}{s}H(p_{1|0})+\frac{p_{1|0}}{s}H(s-p_{1|0})}{\sqrt{\frac{s-p_{1|0}}{s} \cdot \frac{p_{1|0}}{s}}}
\\
=\sqrt{\frac{s-p_{1|0}}{p_{1|0}}}H(p_{1|0})+\sqrt{\frac{p_{1|0}}{s-p_{1|0}}}H(s-p_{1|0})
\end{split} 
\end{equation}
One can check that for smaller $s \in (0,1)$, i.e in case (\ref{A}), for a given fixed $s$ when $p_{1|0}$ tends to interval bounds 0 or to $s$, the quotient $Q_{\sigma}^{M,s}(p_{1|0})$ tends to 0, i.e.:            
\begin{equation}
\lim\limits_{p_{1|0} \to 0^{+}} Q_{\sigma}^{M,s} (p_{1|0})=\lim\limits_{p_{1|0} \to s^{-}} Q_{\sigma}^{M,s} (p_{1|0})=0 .
\end{equation}
By the form of (\ref{Qsigma}) and symmetry property (\ref{Q_M_symmetry}) it is clear that the quotient $Q_{\sigma}^{M,s}(p_{1|0})$ reaches the maximum in the symmetry point $p_{1|0}=\frac{s}{2}$ and it is equal to
\begin{equation}\label{Qsigma_sym}
Q_{\sigma}^{M,s}(\frac{s}{2})=2H(\frac{s}{2}).
\end{equation}
One can check that in the case B) i.e. for $s \in (1,2)$ for a given fixed $s$ when $p_{1|0}$ tends to $s-1$ or to 1 the quotient $Q_{\sigma}^{M,s}(p_{1|0})$ tends to $\frac{H(s-1)}{\sqrt{(s-1)}}$, i.e.: 
\begin{equation}\label{Qlim}
\lim\limits_{p_{1|0} \to (s-1)^{+}} Q_{\sigma}^{M,s} (p_{1|0})=\lim\limits_{p_{1|0} \to 1^{-}} Q_{\sigma}^{M,s} (p_{1|0})=\frac{H(s-1)}{\sqrt{s-1}} .
\end{equation}
Thus, we have for $s \in (1,2)$
\begin{equation}
\frac{H(s-1)}{\sqrt{(s-1)}} \leq Q_{\sigma}^{M,s}(p_{1|0}) \leq 2H(\frac{s}{2}).
\end{equation}
Thus, we obtained an interesting estimation of Information Transmission Rate $ITR$ by the level of fluctuation $\sigma$:
\begin{equation}
\frac{H(s-1)}{\sqrt{s-1}} \sigma^{M}_{s}(p_{1|0})\leq ITR^{M,s}(p_{1|0}) \leq 2H(\frac{s}{2}) \sigma^{M}_{s}(p_{1|0}).
\end{equation}
\begin{figure}
	\centering
	\includegraphics[width=12 cm]{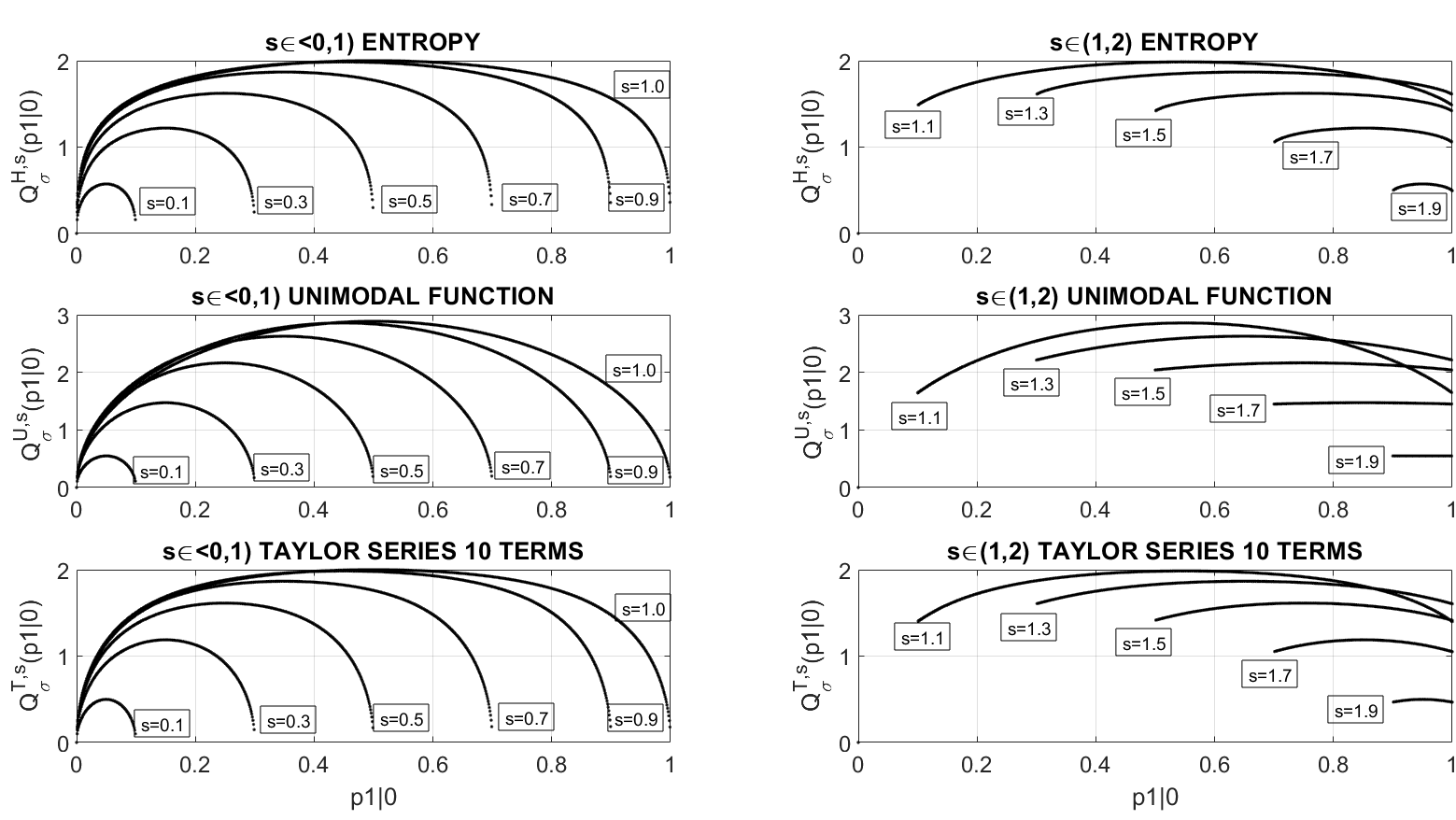}
	\caption{The quotient $\frac{ITR}{\sigma}$ as a function of the transition probability $p_{1|0}$ for chosen values of the jumping parameter $s$: A) For parameters $0 \leq s \leq 1$ due to (\ref{Q_M}) the range of $p_{1|0}$ is $[0,s]$ and B) for $1<s<2$ according to (\ref{Q_M_symmetry}) the range of $p_{1|0}$ is $1-s \leq p_{1|0} \leq 1$. The courses of the quotients $Q_{\sigma}^{H,s}(p_{1|0}),Q_{\sigma}^{U,s} (p_{1|0}),Q_{\sigma}^{T,s}(p_{1|0})$ for Shannon form, unimodal map, Taylor series being applied as $H$ in formula (\ref{entropy_Markov}) are presented.} \label{res_sd}
\end{figure}  
\par
The typical runnings of $Q_{\sigma}^{M,s}(p_{1|0})$ for some values of the parameter, $s$ are shown in Figure \ref{res_sd}. Column A is devoted to lower values of the jumping parameter $0 \leq s \leq 1$, while column B presents the $Q_{\sigma}^{M,s}$ courses for higher values of the jumping parameter $1<s<2$. Observe, that for  $1<s<2$ the curves intersect contrary to the case $0 \leq s \leq 1$. This is mostly since the limiting value (\ref{Qlim}) is not a monotonic function of $s$ while the maximal value (\ref{Qsigma_sym}) is already monotonic.
\par
Note, that for the approximation of entropy $H$ by polynomials, specifically by unimodal map $U$ and by Taylor series $T$, the corresponding quotients $Q_{\sigma}^{U,s}$ B, $Q_{\sigma}^{T,s}$ behave similarly as for the Shannon form of $H$ (see Figure \ref{res_sd}).
\subsubsection{Relation between Information Transmission Rate $ITR$ of Markov process and its Variation}\label{itr_var}
To find how the Variation of trajectories of Markov Information Source affects Information Transmission Rate one should consider now a modified quotient
\begin{equation}\label{qs_itr_var}
Q_{\sigma}^{\bf{M},s}(p_{1|0})=\frac{ITR^{\bf{M}}(p_{1|0})}{V(p_{1|0})}=\frac{ITR^{\bf{M}}(p_{1|0})}{P_{eq}(0) \cdot P_{eq}(1)} .
\end{equation}
Substituting (\ref{stacionary_solution}) and (\ref{s}) to (\ref{qs_itr_var}) we obtain
\begin{equation}
Q_{V}^{\bf{M},s}(p_{1|0})=\frac{{\frac{p_{0|1}}{s}H(p_{1|0})+\frac{p_{1|0}}{s}H(p_{0|1})}}{\frac{p_{0|1}}{s} \cdot \frac{p_{1|0}}{s}}=s[\frac{H(p_{1|0})}{p_{1|0}}+\frac{H(s-p_{1|0})}{s-p_{1|0}}].
\end{equation}
First, observe that clearly as in the standard deviation case we have symmetry property around the value $\frac{s}{2}$,  i.e.
\begin{equation}
Q_{V}^{M,s}(p_{1|0})=Q_{V}^{M,s}(s-p_{1|0}) .
\end{equation}
By this symmetry it is clear that $Q_{V}^{M,s}(p_{1|0})$ reaches extremum at the point $p_{1|0}=\frac{s}{2}$ and it is equal to $4H(\frac{s}{2})$.
\par
Observe, that in the case A), i.e. for a given fixed $s \in (0,1)$, for $p_{1|0}$ tending interval bound i.e. to 0 or $s-$  the quotient $Q_{V}^{M,s}(p_{1|0})$, in opposite to $Q_{\sigma}^{T,s}(p_{1|0})$, tends to infinity, i.e.: 
\begin{equation}
\lim\limits_{p_{1|0} \to 0^{+}} Q_{V}^{M,s}(p_{1|0})=\lim\limits_{p_{1|0} \to s^{-}} Q_{V}^{M,s}(p_{1|0})=+ \infty .
\end{equation}
Thus, it is clear that $Q_{V}^{M,s}(p_{1|0})$ reaches a minimum at the point $p_{1|0}=\frac{s}{2}$. 
\par
In the case of B), it turned out that the quotient $Q_{V}^{M,s}(p_{1|0})$ for any fixed $s \in (1,2)$ is bounded both from below and from above. We have:
\begin{equation}
\lim\limits_{p_{1|0} \to (s-1)^{+}} Q_{V}^{M,s}(p_{1|0})=\lim\limits_{p_{1|0} \to 1^{-}} Q_{V}^{M,s}(p_{1|0})=s\frac{H(s-1)}{s-1} .
\end{equation}
Numerical calculations showed that for the parameters $s>s_{0}$ the point $p_{1|0}=\frac{s}{2}$ is a minimum while for $s<s_{0}$ at this point, there is a maximum, where the critical parameter $s_{0}\approx$1.33 can be calculated from the equality:
\begin{equation}
s_{0} \frac{H(s_{0}-1)}{s_{0}-1}=4H(\frac{s_{0}}{2}) .
\end{equation}
\begin{figure}
	\centering
	\includegraphics[width=12 cm]{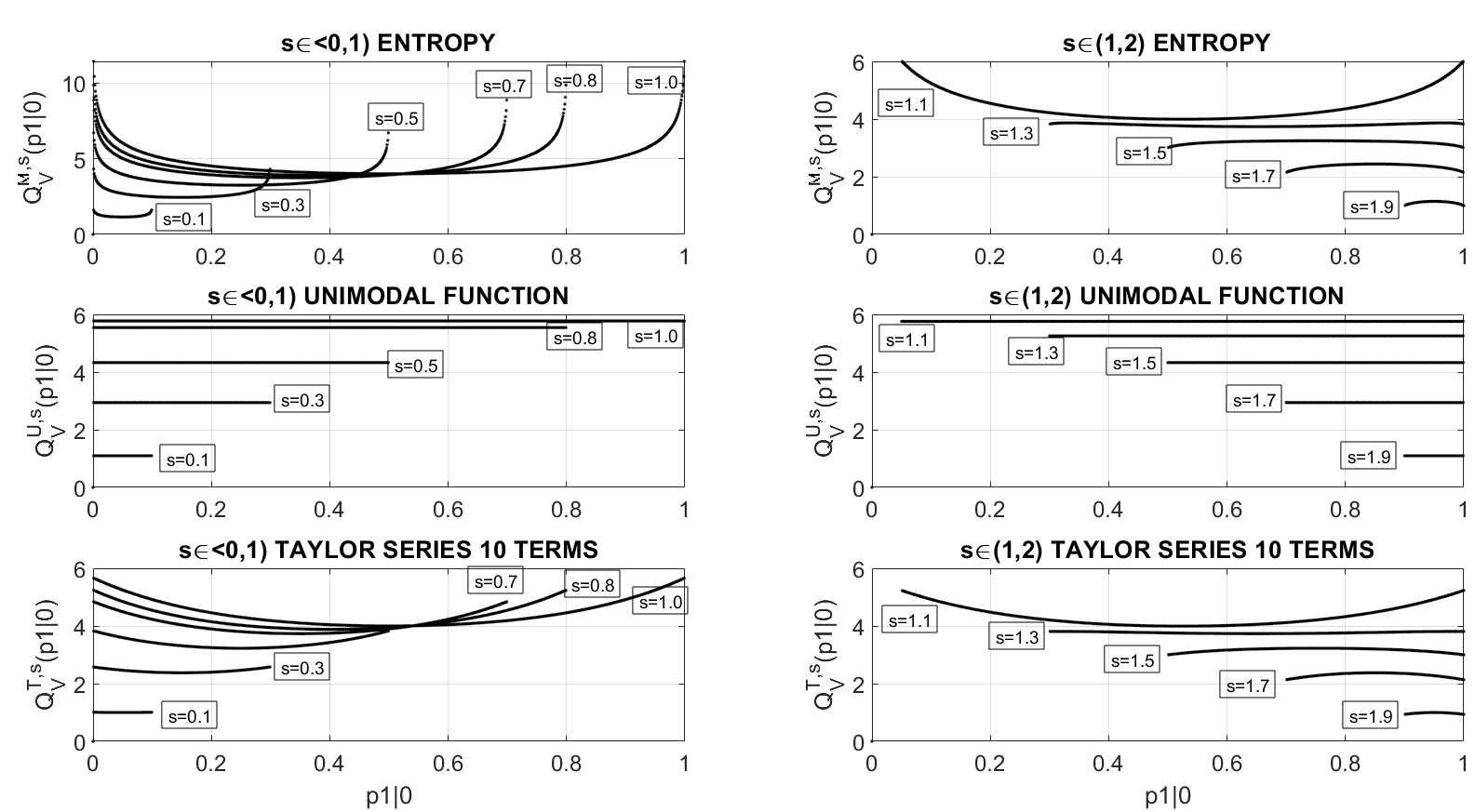}
	\caption{The quotient $\frac{ITR}{V}$ as a function of the initial probability $p_{1|0}$ for the chosen values of the jumping parameter $s$: A) For parameters $0 \leq s \leq 1$ due to (\ref{Q_M}) the range is $0 < p_{1|0} < s$ and B) For parameters $1<s<2$ due to (\ref{Q_M}) the range is $s-1 \leq p_{1|0} \leq 1$.} \label{res_var}
\end{figure}  
The typical running of the $Q_{V}^{M,s}(p_{1|0})$ for some values of the parameter, $s$ is shown in Figure \ref{res_var}. Panel A (left column) is devoted to lower values of the jumping parameter $0 \leq s \leq 1$, while panel B presents graphs of $Q_{V}^{M,s}(p_{1|0})$  for higher values of the jumping parameter $1<s<2$.
\par
It turned out that the approximation of entropy $H$ by polynomials namely by the unimodal map and by Taylor series leads to the completely different behavior of $Q_{V}^{M,s}(p_{1|0})$. Note, that for the approximation of $H$ in (\ref{entropy_Markov}) with the unimodal map the quotient $Q_{V}^{U,s}(p_{1|0})$, for each $s$, is a constant and equal to $4s(2-s)$, while for  the approximation by the Taylor series (10 terms) the quotient $Q_{V}^{T,s}(p_{1|0})$ preserves a similar courses as for $H$ of Shannon form. 
%
%
%
%

\section{Discussion and Conclusions}\label{disscussion}
\par
In this paper, we study relation between the Information Transmission Rate carried out by sequences of bits and these bits fluctuations. These sequences are coming from Information Sources which are modeled by Markov processes. Our results show that the qualitative and quantitative character of the relation between the Information Transmission Rate and signal bits fluctuations strongly depends on the jumping parameter $s$, which we introduced in our previous papers \cite{Pregowska2016,Pregowska2019}. This parameter characterizes the tendency of the process to transition from state to state. In some sense, it describes the variability of the signals. 
\par
It turned out that similarly as in our previous papers when we have studied relation between Information Transmission Rates, spikes correlations, and frequencies of these spikes appearance, the critical value of $s$ is equal to 1 what corresponds to Bernoulli process. For all small $s$ $(s<1)$ the quotient $\frac{ITR}{\sigma}$ can reach 0, while for larger $s$ $(s>1)$ this quotient is always separated from 0.
Specifically, for $1<s<1.7$ the $ITR$ will be always, independently on transition probabilities which forming this $s$, above the level of fluctuations $($i.e. $\sigma<ITR)$. Thus, this shows an interesting fact that for $s$ large enough the information is never completely lost independently on the level of fluctuations. 
\par
On the other hand, for each $0 < s < 2$ the quotient $\frac
{ITR}{\sigma}$ is limited from above by 2 and it is reached for each $s$, for $p_{1|0}=\frac{s}{2}$, i.e. it is reached when $p_{1|0}=p_{0|1}$. This means that, when compare $ITR$ to $\sigma$, the most effective transmission is for symmetric communication channels. Note, that the capacity $C(s)$ of such channels is already equal to
\begin{equation}
C(s)=1-H(\frac{s}{2}).
\end{equation}
It turned out that $\frac{ITR}{\sigma}$ for the approximation of Shannon entropy $H$ by polynomials, specifically by the unimodal map and its Taylor series behaves similarly.   
\par
For better insight, we also referred $ITR$ to Variance. We observed that the behavior of the $\frac{ITR}{V}$ significantly differs from the behavior of $\frac{ITR}{\sigma}$. For each $s<1$ the quotient $\frac{ITR}{V}$ can tend to infinity and it is separated from 0. For $1 < s < 2$ it is limited from above and it never reaches 0 for any $s$. However, it behaves in a more complex way than $\frac{ITR}{\sigma}$ by having even 3 local extreme points, eg. it is visible for $s=1.3$ and $s=1.5$. On the other hand approximations of Shannon entropy $H$ by polynomials like the unimodal map or by its Taylor series, contrary to the case of  $\frac{ITR}{\sigma}$, lead to a significant qualitative difference between the behavior of $\frac{ITR}{V}$.
\par
To summarize, the results obtained show that for Markov information sources, regardless of the level of fluctuation, the level of Information Transmission Rate does not reduce to zero, provided that the transition parameter s is sufficiently large. This means that to get more reliable communication the spike trains should have a higher tendency of transition from the state no spike to spike state and \textit{vice versa}.
\par
The results are presented in the context of signal processing in the brain, due to the fact that information transmission in the brain is in this case a natural and fundamental phenomena. However, our results have, in fact, a general character and can be applied to any communication systems modeled by two states Markov processes.


\bibliography{mybibfile}

\end{document}